\documentclass{optica-article}

\journal{opticajournal} 

\articletype{Research Article}
\usepackage[]{makecell}
\usepackage{multirow}
\usepackage{siunitx}
\usepackage{cite, color, xspace}

\newcommand{\invitro}{\textit{in vitro}\xspace}

\newcommand{\invivo}{\textit{in vivo}\xspace}

\newcommand{\etal}{\textit{et al.\@}\xspace}

\newcommand{\um}{$\mu$m\xspace}

\newcommand{\uM}{\(\mu\)M\xspace}

\graphicspath{{.}{./figures}}

\begin{document}
	\title{High-speed volumetric amplitude-spectrum dynamic optical coherence tomography by neural network with multi-burst scanning}
\author{%
	Yusong Liu,\authormark{1}
	Ibrahim Abd El-Sadek,\authormark{1,2}
	Atsuko Furukawa,\authormark{3}
	Rion Morishita,\authormark{1}
	Satoshi Matsusaka,\authormark{3} and
	Yoshiaki Yasuno\authormark{1,*}
	}

\address{%
	\authormark{1}Computational Optics Group, University of Tsukuba, 1-1-1 Tennodai, Tsukuba, Ibaraki 305-8573, Japan\\
	\authormark{2}Department of Physics, Faculty of Science, Damietta University, New Damietta City 34517, Damietta, Egypt\\
	\authormark{3}Institute of Medicine, University of Tsukuba, Tsukuba, Ibaraki 305-8575, Japan
	}

\email{\authormark{*}yoshiaki.yasuno@cog-labs.org}

\begin{abstract*} 
Dynamic optical coherence tomography (DOCT) enables label-free, three-dimensional (3D) assessment of tissue dynamics.
However, it suffers from long acquisition times because conventional time-spectrum DOCT requires hundreds of repeated OCT frames per location.
Here we present a neural network (NN) framework integrated with a non-uniform-time scanning protocol (multi-burst scan) to accelerate amplitude-spectrum DOCT (AS-DOCT).
Combining 3D convolutional and long-short term memory (LSTM) layers with dual inputs (the temporal OCT sequence and its pseudo-amplitude spectrum), the model generates AS-DOCT images from only 16 frames per location. 
Validated on 29 cancer spheroids, the proposed method resolved distinct functional domain structures with high fidelity (structural similarity index metric (SSIM) > 0.8) and enabled full volumetric AS-DOCT acquisition in 26.2 seconds.
This method will enable high-throughput 3D dynamic tissue screening.
\end{abstract*}

\section{Introduction}
Dynamic optical coherence tomography (DOCT)\cite{josefsberBOE2025, heldt2025BOE, ren2024CB} is a contrast-extension method of optical coherence tomography (OCT)\cite{huang1991Science} that three-dimensionally assesses cellular activity and differentiates tissues based on their functions without using any exogenous agent.
Under the assumption that cellular and intracellular activities cause scatterer motion and subsequent OCT signal fluctuations, DOCT extracts the temporal fluctuations of the OCT signal as a contrast source for assessing scatterer motion and imaging tissue dynamics.
DOCT imaging consists of two steps: sequential acquisition of multiple repeated OCT frames at a single tissue location and extraction of dynamic information from the fluctuations using statistical analysis.
Several DOCT algorithms have been proposed for assessing the properties of scatterer motion, including time-variance-based or standard-deviation-based analysis\cite{apelian2016BOE, Ibrahim2020BOE, Ibrahim2021BOE, morishitaBOE2026}, temporal auto-correlation analysis\cite{oldenburg2015optica, Ibrahim2020BOE, Ibrahim2021BOE}, and time-spectrum analysis \cite{scholler2020Light, HMLeung2020BOE, munterOL2020,munter2021BOE, KYChen2024BOE, Swanson2026BOE}.
These DOCT algorithms have been successfully used to evaluate several properties of scatterer motion, including the occupancy of moving scatterers, the magnitude of scatterer motion, and the speed or frequency of scatterer motion.

Despite the success of DOCT, it requires acquiring a large number of OCT frames (tens to hundreds) at each tissue location, resulting in long acquisition times, especially for three-dimensional (3D) imaging.
This long acquisition time limits further applications of DOCT, such as high-throughput imaging and \invivo imaging.
Although there are a few examples of \invivo DOCT, they require physical fixation for human subjects \cite{YuGuo2026BOE} or proper anesthesia for animals \cite{CYBao2026arXiv}.
Specifically, for variance- or standard-deviation-based DOCT computations, Abd El-Sadek \etal\cite{Ibrahim2021BOE} and Morishita \etal\cite{morishitaBOE2026} used 32 OCT frames per location, whereas Apelian \etal\cite{apelian2016BOE} used 100 frames per location.
For temporal auto-correlation-based DOCT computations, Abd El-Sadek \etal\cite{Ibrahim2021BOE} used 32 frames per location, and Oldenburg \etal\cite{oldenburg2015optica} used 100 frames per location.
For time-spectrum-based DOCT computations, M\"{u}nter \etal\cite{munterOL2020} used 100 to 150 frames per location, and Scholler \etal\cite{scholler2020Light} used 500 frames per location.

To reduce the number of OCT frames and decrease the acquisition time of DOCT, we previously trained a neural network (NN) to generate a time-variance-based DOCT image (i.e., logarithmic intensity variance, LIV\cite{Ibrahim2021BOE}) from only 4 OCT frames instead of the conventional 32 frames.
This method reduced the volumetric acquisition time of LIV from 1 min to 6 s\cite{liu2024BOE}.
However, LIV only represents the magnitude of the signal fluctuation and does not capture speed- or frequency-related information.
To make DOCT measurements comprehensive, a contrast sensitive to speed or temporal frequency is additionally required.

Amplitude-spectrum DOCT (AS-DOCT) \cite{munterOL2020,munter2021BOE, KYChen2024BOE, Swanson2026BOE} is a type of time-spectrum DOCT sensitive to specific bands of the OCT signal's time spectrum.
AS-DOCT typically requires a large number of frames per location (around 100 to 150 frames), and it makes an NN-based reduction in the required frame count highly beneficial.

Although our previous NN method successfully generated LIV images from a small number of OCT frames \cite{liu2024BOE}, it cannot be directly applied to frequency- or speed-sensitive DOCT generation.
The previous NN primarily utilized two-dimensional spatial convolutional layers, which cannot properly handle the time-order information of OCT frames.
Because frequency- or speed-sensitive DOCT computation heavily depends on the time-order information of OCT frames, the previous architecture is unsuitable for this purpose.
In addition, the input OCT frames for LIV generation were acquired with uniform time intervals, preventing the effective transmission of a wide frequency range at a high signal-to-noise ratio (SNR).
Heldt \etal used a neural gas method to enhance the image contrast of AS-DOCT computed from the time spectrum of a small number of OCT frames \cite{heldt2025BOE2}, but their approach suffers from high speckle noise.

To adapt the NN method with a small number of frames for frequency- or speed-sensitive DOCT image generation, we addressed these two issues.
First, we propose a new NN architecture using three-dimensional (3D) convolutional layers combined with a long short-term memory (LSTM) module \cite{hochreiter1997Nc} to extract time-sequential information from the OCT sequence, where 3D refers to 2D spatial dimensions plus time.
Second, we propose a new scanning protocol that captures a small number of OCT frames with non-uniform time separations, ensuring that the NN input effectively covers a wide range of frequency or speed information.
We hypothesized that combining the new NN architecture with the new OCT scanning protocol can effectively extract temporal sequence information and reconstruct frequency- or speed-based DOCT.
Specifically, our proposed method generates frequency-sensitive DOCT (AS-DOCT) from only 16 frames, enabling volumetric AS-DOCT measurement within 26 s.
The proposed method was validated using three types of \invitro cancer spheroids under various treatment and cultivation conditions.
NN performance was confirmed via subjective and objective evaluations of image contrast.

\section{Principle}
\subsection{Amplitude-spectral DOCT}\label{Sec:ConventionalASDOCT}
AS-DOCT is a type of DOCT sensitive to the frequency of cellular motility.
An AS-DOCT image consists of three channels corresponding to three frequency bands: low, middle, and high frequency.
This method was originally proposed by the University of L\"{u}beck \cite{munterOL2020} and is typically computed from 100 to 150 OCT frames per location in conventional implementations.
Specifically, a Fourier transform is performed on the temporal OCT amplitude signal to obtain the frequency spectrum.
The absolute value of the frequency spectrum (the amplitude spectrum) is then split into low-, middle-, and high-frequency bands, which correspond to the low, middle, and high frequencies of tissue dynamics, respectively.
Finally, the amplitude spectrum of each frequency band is integrated and logarithmized.

We aim to generate these low-, middle-, and high-frequency channel images of AS-DOCT using our newly designed NN (Section \ref{Sec:NNArchitecture}) from only 16 OCT frames acquired via a non-uniform time-scanning protocol (Section \ref{Sec:BurstScan}).

\subsection{Neural network architecture}\label{Sec:NNArchitecture}
\begin{figure}
	\centering\includegraphics[width=13cm]{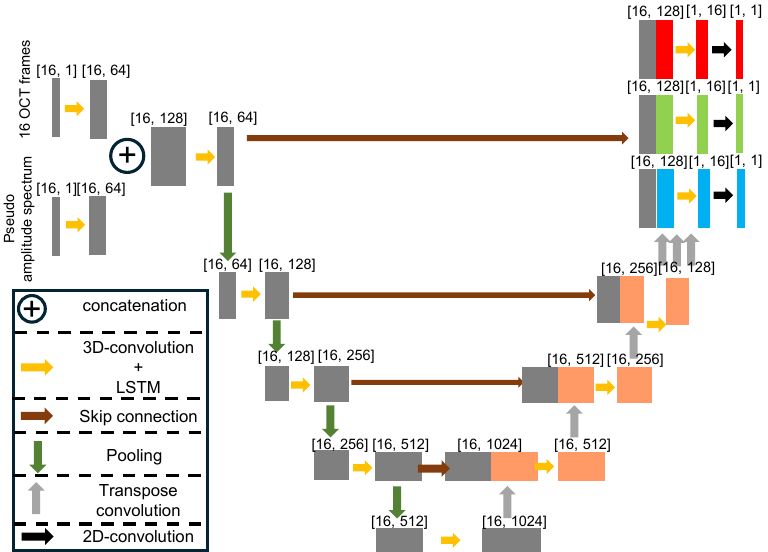}
	\caption{Our proposed NN architecture for extracting sequential information from 16 OCT frames to generate AS-DOCT.
		The NN architecture accepts an OCT time sequence and a pseudo-amplitude spectrum as dual inputs.
		3D convolution and LSTM layers are used to extract the sequential information of the OCT time sequence at different spatial resolutions.
		The decoder branches into three outputs at the final layer to generate low-, middle-, and high-frequency AS-DOCT images.}
	\label{Fig:NN}
\end{figure}
The proposed NN architecture is shown in Fig.\@ \ref{Fig:NN}, where the input consists of two parts.
One is a cross-sectional 16-frame OCT time sequence with non-uniform time spacing, and the other is the pseudo-amplitude spectrum, which is the amplitude of the fast Fourier transform (FFT) spectrum of the input 16 OCT frames.
Here, ``pseudo'' is used because a standard FFT was directly applied to the non-uniformly spaced temporal sequence.
This pseudo-amplitude spectrum input provides supplemental frequency information.
Each input is five-dimensional, with the shape [batch size, time or pseudo-spectrum points, height, width, feature channels].
The NN outputs three images corresponding to the low-, middle-, and high-frequency channels of AS-DOCT.
Each output is a three-dimensional image with the shape [batch size, height, width].
The two numbers in brackets in Fig.\@ \ref{Fig:NN} represent the number of time or pseudo-spectrum points and the number of feature channels, respectively.

This NN architecture uses two core modules to extract temporal information.
One is the 3D convolutional layer, which convolves sequential images along both spatial and temporal axes.
Temporal convolution enables the extraction of order-sensitive temporal features among neighboring frames.
Specifically, the 3D convolutional layer consists of 3D convolution with a 2 $\times$ 2 $\times$ 2 kernel, rectified linear unit (ReLU) activation, and layer normalization.
The other core module is the LSTM layer, which captures long-range temporal dependencies within the image sequence.
It enables the extraction of temporal features between frames with long time intervals.
The LSTM layer is implemented using the Conv2dLSTM( ) function in Keras, with a kernel size of 2 $\times$ 2.
The combination of 3D convolution and the LSTM layer (Conv3D-LSTM module) extracts sequential information across a wide range of frame separations \cite{manttari2020Accv}.

The NN architecture is a modified U-Net\cite{ronneberger2015UNet} consisting of three parts: an encoder, a decoder, and skip connection layers.
The encoder extracts features, the decoder reconstructs the image, and the skip connections connect the encoder and decoder to preserve multi-scale features.
In the encoder, feature representations extracted from the OCT image and pseudo-amplitude spectrum are concatenated along the feature channel dimension.
Then, each pooling layer decreases the spatial image size to 1/4 (1/2 in height $\times$ 1/2 in width) using a combination of max pooling and average pooling.
Throughout the encoder, pooling is performed four times, reducing the spatial dimensions to 1/16 of the original input in both height and width.
This cascading spatial reduction allows the Conv3D-LSTM module to extract information from OCT frames at multiple spatial resolutions.
The feature maps then pass from the encoder into the decoder.
The decoder increases the spatial resolution using transposed convolutional layers.
Each transposed convolutional layer quadruples the spatial image area (doubling both height and width).
Multi-scale features are preserved by concatenating encoder and decoder representations via skip connection layers.
After four transposed convolutional layers, the spatial resolution matches the original input.

At the final layer of the decoder, the architecture branches into three output heads to separately reconstruct the low-, middle-, and high-frequency AS-DOCT images.
Losses from the low-, middle-, and high-frequency AS-DOCT images are computed separately and summed to form the final loss during model training, as described in Section \ref{Sec:TrainingDetail}.

\subsection{Multi-burst scanning protocol}\label{Sec:BurstScan}
\begin{figure}
	\centering\includegraphics[width=10cm]{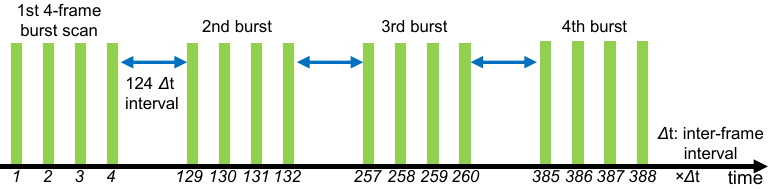}
	\caption{Our proposed multi-burst scanning protocol for acquiring a 16-frame OCT time sequence with both short and long inter-frame intervals.
		Green bars indicate acquisition time points.
	The protocol consists of four burst scans.
	Each burst scan continuously captures four OCT frames.
		An idle interval (equivalent to the acquisition time of 124 frames) is maintained between adjacent burst scans.
	}
	\label{Fig:Burst}
\end{figure}
As described in Section \ref{Sec:NNArchitecture}, the function of the NN is to extract hidden features from a 16-frame OCT time sequence.
To facilitate the extraction of various frequency components, we designed a non-uniform time-scanning protocol (Fig.\@ \ref{Fig:Burst}) so that the 16-frame OCT sequence effectively conveys a wide frequency range information.

This scanning protocol captures 16 OCT frames with both short and long inter-frame intervals.
It first continuously captures 4 frames within a short duration of 4 $\Delta t$, referred to as a ``burst,'' where $\Delta t$ represents the single-frame acquisition time (12.8 ms in our specific implementation, as described in Section \ref{Sec:OCTDevice}).
The burst provides high-frequency information with a high SNR. 
Four bursts are repeated while maintaining an idle interval (equivalent to 124 $\Delta t$) between adjacent bursts.
This idle time intentionally extends the total acquisition duration of the frame sequence to effectively encode low-frequency information with a high SNR.
This protocol is denoted as the ``multi-burst scanning protocol.''

Furthermore, in volumetric imaging implementations, the idle interval is utilized for burst scanning of other spatial locations, enabling volumetric DOCT acquisition without dead time.
Details of this high-speed volumetric DOCT are demonstrated in Section \ref{Sec:HighSpeedAS}.

\section{Implementation and evaluation protocol}
\subsection{Samples}\label{Sec:Samples}
We used a total of 218 cancer spheroids for NN training and evaluation.
They consisted of three types of cancer spheroids: 155 breast cancer (MCF-7 cell line) spheroids formed by seeding 1,000 cells, 58 colon cancer (HT-29 cell line) spheroids formed by seeding 1,000 cells, and 5 breast cancer (MCF-7 cell line) spheroids formed by seeding 5,000 cells.
Details of sample preparation for the 1,000-cell spheroids can be found in \cite{ibrahim2023SciRep,ibrahim2024SciRep}.
The 5,000-cell spheroids were cultivated using the same protocol as the 1,000-cell spheroids, except 5,000 cells were seeded.
All 5,000-cell spheroids were cultivated for 11 days.

The NN training process used only 1,000-cell spheroids, including 143 MCF-7 and 46 HT-29 spheroids.
To increase sample variation for training, spheroids were treated with drugs under various conditions.
MCF-7 spheroids were treated with paclitaxel (PTX) at 0.1-\uM, 1-\uM, and 10-\uM concentrations; tamoxifen (TAM) at 0.1-\uM, 1-\uM, and 10-\uM concentrations; doxorubicin (DOX) at 0.1-\uM, 1-\uM, and 10-\uM concentrations; or left untreated.
Treatment durations varied across 1, 3, and 6 days.
HT-29 spheroids were subjected to treatments using SN-38 at concentrations of 0, 0.1, 1, and 10 \uM for durations of 1, 3, or 6 days, where 0 \uM indicates no treatment.

The remaining samples were reserved for NN evaluation.
These consisted of 1,000-cell MCF-7 spheroids (n = 12), 1,000-cell HT-29 spheroids (n = 12), and 5,000-cell MCF-7 spheroids (n = 5).
All 5,000-cell spheroids were untreated.
The 12 MCF-7 evaluation spheroids were prepared under 12 distinct conditions, defined by four treatment groups times three treatment durations.
The treatment conditions included untreated controls and PTX treatments at 0.1 \uM, 1 \uM, and 10 \uM.
Treatment durations were 1, 3, and 6 days.
Similarly, the 12 HT-29 evaluation spheroids were prepared under 12 distinct conditions combining untreated controls or SN-38 treatments (0.1 \uM, 1 \uM, 10 \uM) across durations of 1, 3, and 6 days.
All 5,000-cell seeded MCF-7 spheroids remained untreated.

\subsection{OCT device for data acquisition}\label{Sec:OCTDevice}
A Jones-matrix swept-source OCT (SS-OCT) system was utilized for acquiring data.
Details of this OCT system can be found in \cite{yasunoIEEE2023,li2017BOE,miyazawa2019BOE}.
Briefly, the center wavelength of the probe beam is 1.3 \um, and the A-line rate is 50 kHz.
The axial and lateral resolutions are 14 \um (in tissue) and 18.1 \um, respectively.
Although this system possesses polarization sensitivity, only polarization-insensitive intensity images (obtained by intensity-averaging of four polarization channels) were used in this study.

\subsection{Implementation method for training the NN model}
\subsubsection{Ground truth and NN input}\label{Sec:GroundTruthandInput}
To compute ground truth data for NN training, we implemented conventional AS-DOCT as described in Section \ref{Sec:ConventionalASDOCT} on our OCT system.
For each spheroid, continuous 512 OCT frames were captured at a single location over a lateral field of view of 1 mm.
Each frame comprised 512 A-lines, corresponding to a frame acquisition time of 12.8 ms and a Nyquist frequency of 40 Hz.
Ground truth AS-DOCT was computed from the full 512-frame sequence using the conventional method.
An FFT yielded an amplitude spectrum with 256 frequency bins.
Similar to Ref.\@\cite{munterOL2020}, the spectrum was split into three frequency bands: low frequency [0--0.5 Hz], middle frequency [0.5--5 Hz], and high frequency [5--25 Hz].
The spectrum in each frequency band was integrated using the numpy.trapz( ) function, logarithmized using numpy.log10( ), and designated as ground truth.

For the NN input, 16 frames were extracted from the same 512-frame sequence used for ground truth generation.
To match the multi-burst scanning protocol described in Section \ref{Sec:BurstScan}, the extracted frames were the 1st--4th, 129th--132nd, 257th--260th, and 385th--388th frames of the 512-frame sequence.
The short intra-burst frame interval was 12.8 ms, while the long inter-burst interval was 1.59 s (that is equivalent to 12.8 ms/frame $\times$ 124 frames).

\subsubsection{Datasets for the training process}
As described in Section \ref{Sec:Samples}, 143 1,000-cell MCF-7 spheroids and 46 1,000-cell HT-29 spheroids were used for NN training.
These samples were randomly assigned into training and validation sets at a 4:1 ratio.
Specifically, the training set consisted of 118 MCF-7 spheroids and 34 HT-29 spheroids, while the validation set comprised 25 MCF-7 spheroids and 12 HT-29 spheroids.
Training samples were used to update model parameters, and validation samples were used to monitor training performance.

The cross-sectional 16-frame sequence for input and the corresponding ground truth were obtained as described in Section \ref{Sec:GroundTruthandInput} and paired.
Image pairs were cropped into 64 $\times$ 64-pixel patches with an overlap of 32 pixels in both depth and fast-scan directions.
Only patches containing the sample region were used for training, where sample regions were manually delineated using ImageJ.
Consequently, the training and validation datasets contained 5,362 and 1,350 patch pairs, respectively.

\subsubsection{NN training details}\label{Sec:TrainingDetail}
During training, the batch size was set to 1.
Before each epoch, the training dataset was shuffled.
Mean squared error (MSE) served as the loss function.
We first calculated the MSE for each frequency channel between the model output and ground truth.
The average MSE across all three channels was then computed as the total loss.
The Adam optimizer was used with learning rate decay: $\text{learning rate} = (0.0005 / \text{epoch}) + 0.0001$.
Training was stopped when validation loss failed to decrease for 5 consecutive epochs, and the checkpoint with the lowest validation loss was selected for inference.

The NN was implemented using TensorFlow-GPU version 2.6 and trained on an NVIDIA RTX 4090 GPU with 24 GB of VRAM.
Model training took 25.5 hours over 21 epochs.

\subsection{Evaluation method of the NN model}
The proposed NN method was evaluated via two studies.
Study 1 evaluated NN performance by comparing contrast similarity between 16-frame NN-generated images and 512-frame conventional AS-DOCT images.
Study 2 performed an ablation analysis to confirm the necessity of the multi-burst scanning protocol and the dual-input strategy.

\subsubsection{Image types for evaluation}\label{Sec:ImageTypes}
Evaluation was conducted across four image types: low-, middle-, and high-frequency band images, and composite pseudo-color RGB AS-DOCT images.
Unlike the training phase, three-band images during evaluation were generated directly from full-size cross-sectional OCT images rather than small patches.
Thus, full cross-sectional AS-DOCT images were produced in a single inference step.
Ground truth images were derived from 512-frame sequences as described in Section \ref{Sec:GroundTruthandInput}.

Composite pseudo-color RGB images were formed by combining the three single-channel outputs to visualize domain structures and differentiate tissue regions.
We adopted an approach similar to \cite{munter2021BOE} for RGB image generation.
Each channel was normalized to the range [0, 1], after which high-, middle-, and low-frequency images were assigned to the red, green, and blue channels, respectively.

\subsubsection{Study 1: Protocol for evaluating NN performance}
\label{Sec:EvaluationProtocolForSimilarity}
Similarity between ground truth and NN-generated images was evaluated on three spheroid types: 1,000-cell MCF-7 spheroids (n = 12), 1,000-cell HT-29 spheroids (n = 12), and 5,000-cell MCF-7 spheroids (n = 5), as described in Section \ref{Sec:Samples}.
Notably, because the NN model was trained exclusively on 1,000-cell spheroids, the five 5,000-cell spheroids evaluated the model's generalization capabilities.

Ground truth and NN-generated AS-DOCT images were derived from identical data sequences.
Specifically, the 16-frame input sequence was extracted from the 512-frame sequence used to compute the ground truth.

Image contrast similarity was evaluated both subjectively and objectively.
First, qualitative visual features were compared between NN outputs and ground truth.
Because cancer spheroids exhibit distinct structural domains (such as proliferating and necrotic zones \cite{nayak2023Cancer}), subjective assessment focused on domain visualization.
Second, the structural similarity index (SSIM) within the sample region was computed to quantify contrast fidelity.
SSIM values were calculated using the skimage.metrics.structural\_similarity( ) function in scikit-image (version 0.25).
The data range parameter was set to the maximum value of the ground truth image.
The function evaluated local SSIM using a 7 $\times$ 7-pixel sliding window to produce a spatial SSIM map.
This map was multiplied by a segmentation mask (manually segmented using ImageJ) to compute average SSIM strictly within the sample boundaries.

Quantitative metrics derived from DOCT images are commonly used to assess tissue dynamics \cite{ibrahim2023SciRep, ibrahim2024SciRep, ElSadek2026SciRep}.
Here, we evaluated agreement between quantitative metrics extracted from NN-AS-DOCT and ground-truth AS-DOCT images.
Specifically, mean AS-DOCT values for each frequency band were measured across three regions of interest (ROIs): core, intermediate, and periphery ROIs.
Untreated 5,000-cell MCF-7 spheroids (n = 5) were used for this evaluation.

\subsubsection{Study 2: Ablation study protocol}
In the ablation study, we investigated the necessity and importance of both the multi-burst scanning protocol and the dual-input strategy.
For comparison, two additional NN models were trained.
The first model accepted only 16 OCT frames and omitted the pseudo-amplitude spectrum input.
This variant is designated as the ``single-input burst NN model,'' whereas the primary architecture is called the ``dual-input burst NN model''.
The second modified model used dual inputs, but the 16 OCT frames were extracted with uniform temporal spacing (the 1st, 27th, 53rd, ..., and 391st frames of the 512-frame sequence).
The inter-frame interval was 332.8 ms and the total time window was 4.99 s, matching the time window of the multi-burst scan protocol (4.95 s) to eliminate total-duration bias.
This model is designated as the ``dual-input uniform NN model''.

Data from the 5,000-cell spheroids were used for this study.
Outputs from all models were compared subjectively by visual inspection and objectively using SSIM metrics.

\section{Results}
\subsection{Study 1: Results of NN performance evaluation}
\subsubsection{Image contrast of NN-AS-DOCT}\label{Sec:ResultImageContrast}
\begin{figure}
	\centering\includegraphics[width=13cm]{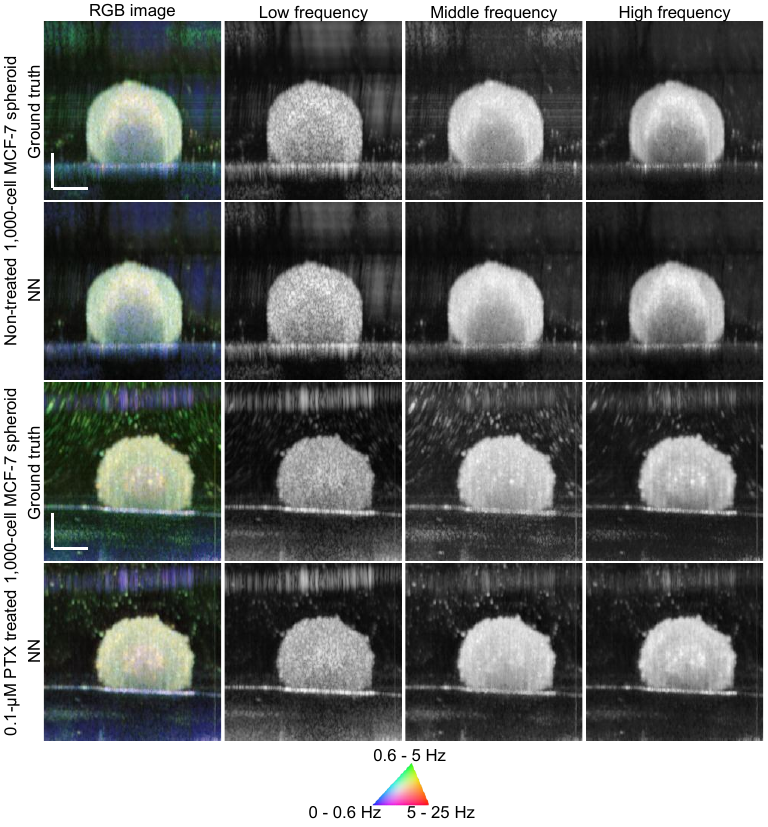}
	\caption{%
		Ground truth and NN-generated cross-sectional AS-DOCT images of 1,000-cell MCF-7 spheroids.
		The top row shows an untreated spheroid, and the bottom row shows a spheroid treated with 0.1-\uM PTX.
		Both ground truth and NN-AS-DOCT reveal a three-domain structure in the untreated spheroid and a two-domain structure in the drug-treated spheroid.
		Scale bar: 200 \um.
		See supplementary Fig.\@ S1 for the examples of other spheroids.
	}
	\label{Fig:MCF1000}
\end{figure}
\begin{figure}
	\centering\includegraphics[width=13cm]{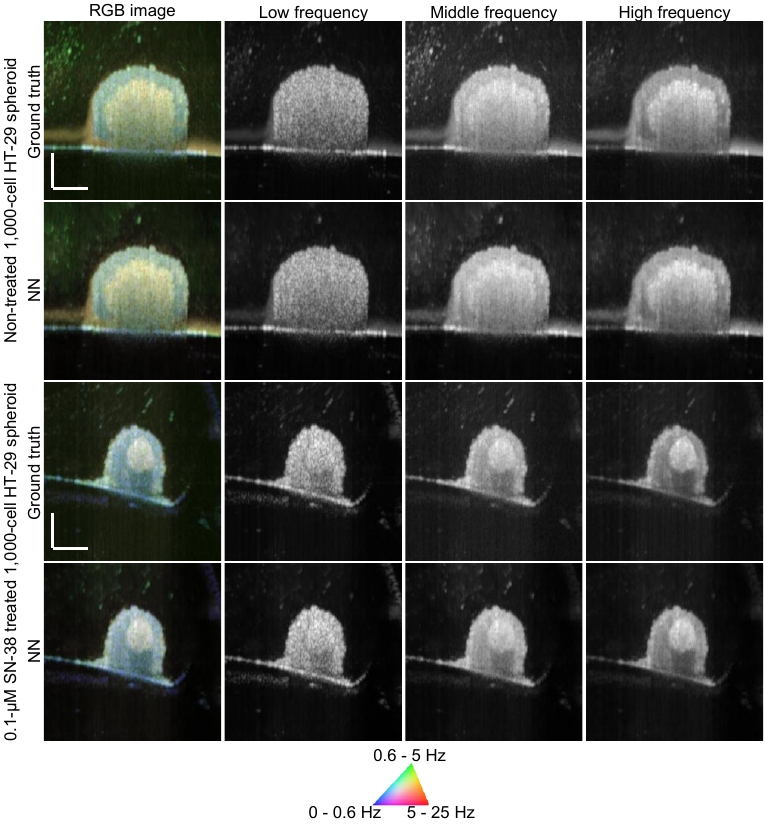}
	\caption{Ground truth and NN-generated cross-sectional AS-DOCT images of 1,000-cell HT-29 spheroids.
		The top row shows an untreated spheroid, and the bottom row shows a spheroid treated with 0.1-\uM SN-38.
		Both ground truth and NN-AS-DOCT reveal similar two-domain structures in both spheroids.
		Scale bar: 200 \um.
		See supplementary Fig.\@ S2 for the examples of other spheroids.
		}
	\label{Fig:HT1000}
\end{figure}
Figure \ref{Fig:MCF1000} shows ground truth and NN-generated AS-DOCT images of untreated and 0.1-\uM PTX-treated 1,000-cell MCF-7 spheroids.
In the untreated spheroid, three structural domains are apparent in RGB composites for both ground truth and NN predictions: blue in the core, green in the periphery, and yellow in the intermediate region.
In the PTX-treated spheroid, two domains (core and periphery) are resolved by both methods.
Figure \ref{Fig:HT1000} shows two representative 1,000-cell HT-29 spheroids.
One sample is untreated (rows 1 and 2), and the other was treated with 0.1-\uM SN-38 (rows 3 and 4).
In both spheroids, two distinct domains are visualized in ground truth and NN-generated AS-DOCT images.

Mean SSIM values within the spheroid region for low-, middle-, and high-frequency bands across 12 MCF-7 spheroids and 12 HT-29 spheroids are summarized in Table \ref{tab:SSIMOfSpheroids}.
Values are expressed as mean $\pm$ standard deviation.
For low- and high-frequency channels, the NN achieved mean SSIMs exceeding 0.9, indicating high agreement with ground truth.
For middle-frequency channels, mean SSIMs were 0.860 for HT-29 spheroids and 0.871 for MCF-7 spheroids.
Although slightly lower than low and high frequencies, middle-frequency SSIMs remained high ($>0.8$).
Notably, the 5,000-cell spheroids also exhibited high SSIM values, confirming that the NN generalizes well to sample sizes not encountered during training.
\begin{table}[]
	\caption{SSIM computed between ground truth and NN predictions across three cancer spheroid types.
		Values represent mean SSIM $\pm$ standard deviation.
		In all three frequency channels across all sample types, SSIM values exceed 0.8.
	}
	\centering
	\setlength{\tabcolsep}{2pt} 
	\begin{tabular}{c|c|c|c}
		& Low frequency & Middle frequency & High frequency\\ \hline
		1,000-cell MCF-7 spheroid & 0.970 $\pm$ 0.0133 & 0.871 $\pm$ 0.0265 & 0.928 $\pm$ 0.0146 \\ \hline
		1,000-cell HT-29 spheroid& 0.968 $\pm$ 0.0076 & 0.860 $\pm$ 0.0081 & 0.918 $\pm$ 0.0081 \\ \hline
		5,000-cell MCF-7 spheroid & 0.949 $\pm$ 0.0035 & 0.800 $\pm$ 0.0017 &  0.898 $\pm$ 0.0101 \\
	\end{tabular}
	\label{tab:SSIMOfSpheroids}
\end{table}

\begin{figure}
	\centering\includegraphics[width=13cm]{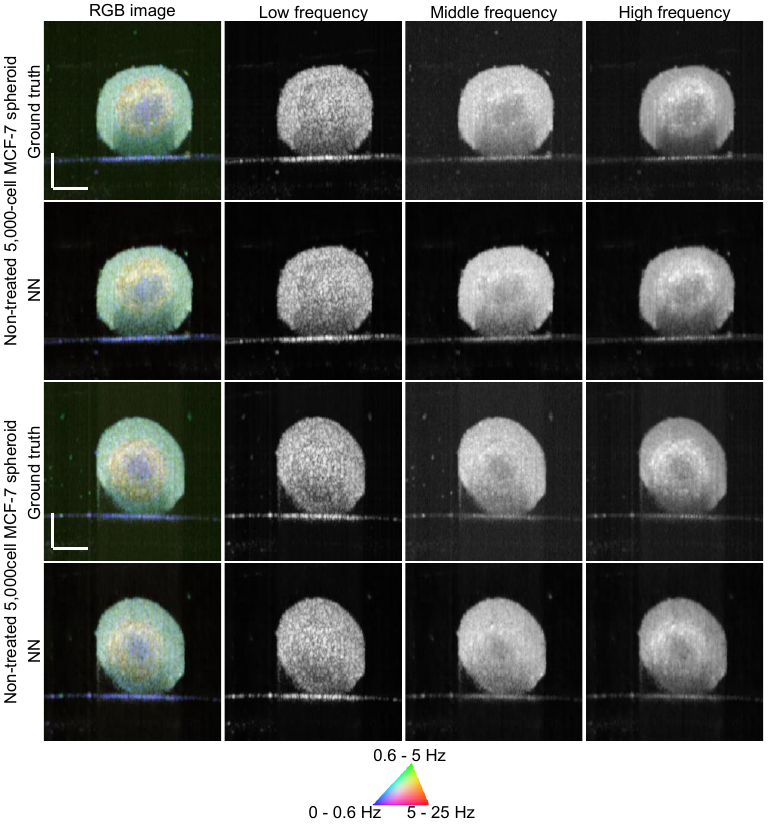}
	\caption{%
		NN-generated and ground truth cross-sectional AS-DOCT images of 5,000-cell MCF-7 spheroids.
		Both NN predictions and ground truth images display matching three-domain structures.
		Scale bar: 200 \um.
		}
	\label{Fig:5000MCF7Burst}
\end{figure}
Figure \ref{Fig:5000MCF7Burst} demonstrates ground truth and NN AS-DOCT images for two 5,000-cell MCF-7 spheroids.
Although spheroids of this size were excluded from the training set, the model produced high-fidelity reconstructions.
Ground truth images show three distinct radial domains from the core to the periphery.
The core exhibits low-frequency-dominant dynamics (blue).
The periphery exhibits middle-frequency-dominant dynamics (green).
The intermediate domain between core and periphery shows middle- to high-frequency-dominant dynamics (yellow).
NN predictions accurately reproduce these dynamic domain patterns.
Mean SSIM values for low-, middle-, and high-frequency channels (ground truth vs. NN) are listed in the final row of Table \ref{tab:SSIMOfSpheroids}.
Although SSIM values are slightly lower than those of 1,000-cell spheroids, all frequency bands maintain values $>0.8$.
It demonstrates model generalization to unseen culture conditions.

The images generated from all other spheroids are presented in supplementary figures (Figs.\@ S1 and S2).

\subsubsection{Mean-AS-DOCT-value analyses}
\label{Sec:AS-DOCT-metrics}

\begin{figure}
	\centering\includegraphics[width=13cm]{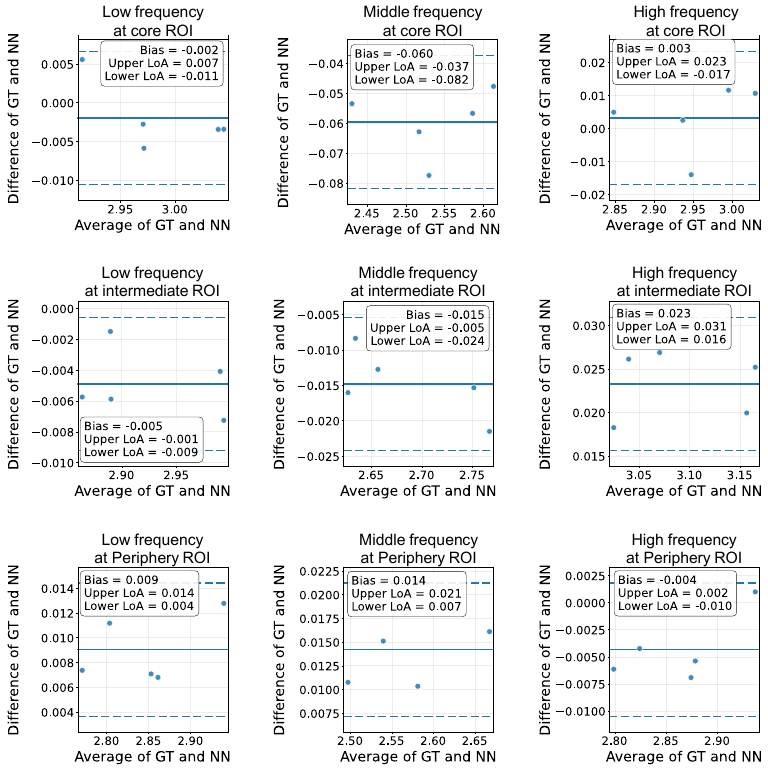}
	\caption{
		Bland-Altman plots comparing ground-truth-derived and NN-derived mean AS-DOCT values.
		The nine plots represent combinations of three regions (core, intermediate, periphery ROIs) and three frequency bands (low, middle, high).
		Most regions exhibit low bias and narrow limits of agreement (LoA), except the core ROI in middle- (bias: -0.06, LoA: 0.045) and high-frequency bands (bias: 0.003, LoA: 0.04).
	}
	\label{Fig:BlandAltman}
\end{figure}
Figure \ref{Fig:BlandAltman} presents Bland-Altman plots of mean AS-DOCT values derived from ground truth versus NN predictions across 5,000-cell MCF-7 spheroids (n = 5).
In these plots, bias (center horizontal line) denotes the mean difference ($d$) between mean AS-DOCT values of GT and NN-derived images.
Lower and upper limits of agreement (LoAs, dashed lines) are defined as bias $\pm$ 1.96 $\times$ SD, where SD is the standard deviation of $d$.

In most cases (excluding the core ROI in middle and high frequencies), ROIs exhibited small bias (around $\pm0.02$) and narrow LoA spans ($<0.02$).
In contrast, the core ROI in the middle-frequency band showed a bias of $-$0.06 and an LoA span of 0.045.
In the high-frequency band, the core ROI showed a low bias of 0.003 but a wide LoA span of 0.04.
These results indicate that NN-derived mean AS-DOCT values agree closely with ground truth across most ROIs and frequency bands, although agreement slightly decreases in the core region for middle and high frequencies.
This limitation is discussed further in Section \ref{Sec:Limitation}.

\subsection{Study 2: Results of the ablation study}
\begin{figure}
	\centering\includegraphics[width=13cm]{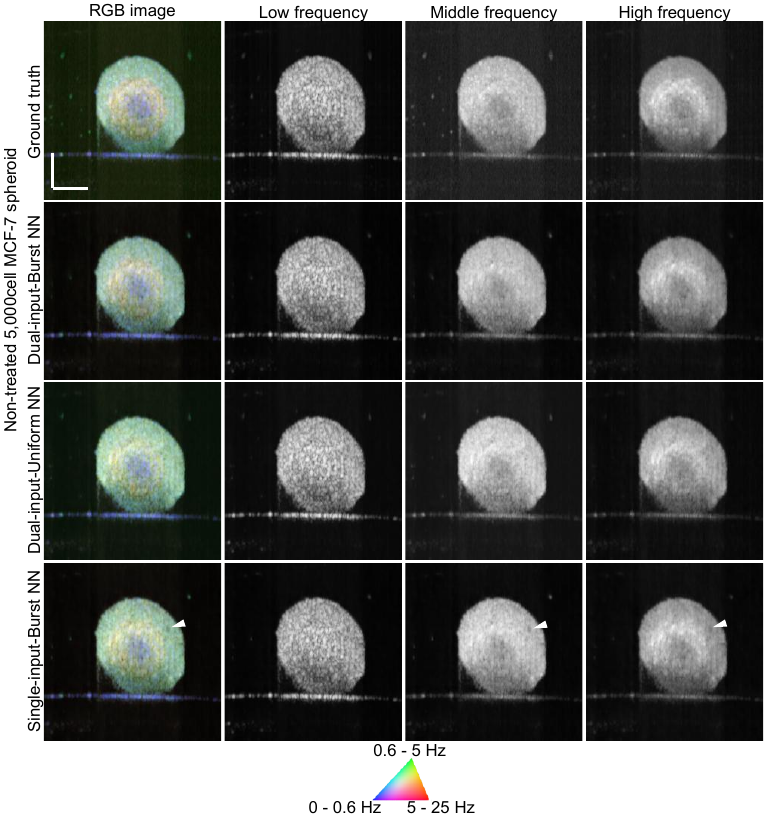}
	\caption{%
		The performance comparison of three NN models.
		Cross-sectional AS-DOCT images from (top row) ground truth, (second row) dual-input burst NN, (third row) dual-input uniform NN, and (bottom row) single-input burst NN  for a 5,000-cell MCF-7 spheroid.
		The dual-input burst NN reproduces the three-domain structure seen in ground truth.
		The dual-input uniform NN produces an artifactual blue boundary between intermediate and peripheral domains.
		The single-input burst NN produces artifactual dark spot (white arrow head).
		Scale bar: 200 \um.}
	\label{Fig:BurstVsUniformScan}
\end{figure}

SSIM values for the dual-input burst NN, single-input burst NN, and dual-input uniform NN are summarized in Table \ref{tab:SSIMBurstvsUniformvsSingleInput}.
Comparing the dual-input burst NN with the single-input burst NN shows that omitting the pseudo-amplitude spectrum reduces SSIM across all three frequency channels.
Visual comparison (Fig.\@ \ref{Fig:BurstVsUniformScan}) demonstrates the single-input burst NN produces artifactual dark spots as indicated by white arrows.
This highlights the utility of dual inputs (combining the OCT sequence with its pseudo-amplitude spectrum).
\begin{table}[]
	\caption{%
		SSIM values calculated against ground truth using datasets from five untreated 5,000-cell MCF-7 spheroids.
	}
	\centering
	\setlength{\tabcolsep}{2pt} 
	\begin{tabular}{c|c|c|c}
		& Dual-input burst NN  &  Dual-input-uniform NN & Single-input burst NN\\
		\hline
		Low frequency & 0.949 $\pm$ 0.0035 & 0.965 $\pm$ 0.0182 & 0.945 $\pm$ 0.004 \\ \hline
		Middle frequency & 0.800 $\pm$ 0.00165 & 0.818 $\pm$ 0.0162 & 0.793 $\pm$ 0.019 \\\hline
		High frequency & 0.898 $\pm$ 0.0101 & 0.881 $\pm$ 0.0102 & 0.886 $\pm$ 0.0122 \\ 
	\end{tabular}
	\label{tab:SSIMBurstvsUniformvsSingleInput}
\end{table}

Comparing the dual-input burst NN with the dual-input uniform NN reveals that uniform sampling yields higher SSIMs for low and middle frequencies, but a lower SSIM for high frequencies.
This result is reasonable because time-sparse uniform sampling cannot capture frequencies above its Nyquist limit (1.5 Hz).
Visual comparison (Fig.\@ \ref{Fig:BurstVsUniformScan}) demonstrates that the boundary between intermediate and peripheral domains is poorly defined in the dual-input uniform model (see the high-frequency-band images).
This degradation stems from reduced contrast in the high-frequency channel, emphasizing the necessity of the multi-burst scanning protocol.

The visual comparison of other spheroids is presented in a supplementary figure (Fig.\@ S3).

\subsection{High-speed volumetric AS-DOCT}
\label{Sec:HighSpeedAS}
The model was further evaluated for 3D volumetric AS-DOCT imaging.
As detailed in Section \ref{Sec:BurstScan}, the multi-burst scanning protocol incorporates an idle interval equal to 124 $\Delta t$ between bursts.
This idle duration was leveraged to scan 31 additional spatial locations (4 frames/location $\times$ 31 locations).
We implemented a volumetric multi-burst scanning protocol covering 128 locations per volume $\times$ 16 frames per location $\times$ 512 A-lines per frame.
Using our SS-OCT system, a full volumetric multi-burst scan takes 26.2 s.
Untreated 5,000-cell MCF-7 spheroids were imaged with this protocol to construct 3D AS-DOCT volumes.

\begin{figure}
	\centering\includegraphics[width=13cm]{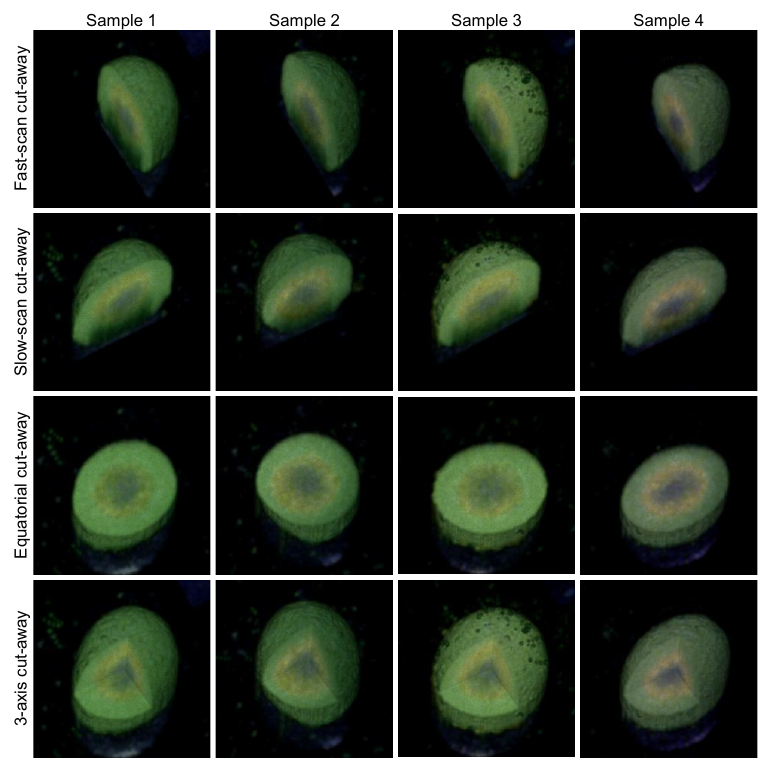}
	\caption{%
		Four volumetric AS-DOCT renderings generated from 16 frames per location using the NN model.
		All samples are untreated 5,000-cell MCF-7 spheroids.
	}
	\label{Fig:VolumeASDOCT}
\end{figure}
Figure \ref{Fig:VolumeASDOCT} shows several types of cut-away volume renderings of RGB AS-DOCT images generated via the Volume Viewer plugin in Fiji/ImageJ.
The three-domain architecture is observable across multiple viewing angles.

Acquisition times for NN-based AS-DOCT and conventional AS-DOCT implementations reported in the literature are summarized in Table \ref{tab:OurMethodVsLubeck}.
Our method (first row) requires the fewest OCT frames per location—approximately one-tenth of conventional methods.
Compared to conventional AS-DOCT on systems of similar speed (rows 2 and 3), our NN-based method reduces acquisition time by a factor of 10 to 20.
Even compared to AS-OCT on ultra-high-speed systems (600 kHz A-line rate, row 4), NN-based AS-DOCT on a standard 50-kHz system achieves comparable volumetric acquisition times.
Thus, the proposed method enables rapid 3D AS-DOCT using standard OCT hardware.
\begin{table}[]
	\caption{%
		Comparison between NN-based AS-DOCT and conventional AS-DOCT implementations.
		Our method uses the fewest OCT frames per location and achieves a rapid volumetric acquisition time of 26.2 s per volume.
		$\dagger$ indicates estimated values assumed by the authors when omitted from original literature.
	}
	\centering
	\setlength{\tabcolsep}{2pt} 
	\begin{tabular}{c|c|c|c|c|c|c}
		& A-line rate 
		& \begin{tabular}{c}Frequency\\range \end{tabular}
		& \begin{tabular}{c}Frames\\/location \end{tabular}
		& \begin{tabular}{c}Locations\end{tabular}
		&  \begin{tabular}{c}Volume\\acquisition time\end{tabular}
		&  \begin{tabular}{c}Time for\\512 locations\end{tabular}\\ 
		\hline
		Ours & 50 kHz & 0 - 25 Hz & 16 & 128 & 26.2s & 1.7 min\\ \hline
		L\"ubeck \cite{munterOL2020}& 30 kHz & 0 - 25 Hz & 100 & $^\dagger$ 128 & 4.6 min & 18.4 min\\ \hline
		L\"ubeck\cite{kohlfaerberBOE2022 }& 100 kHz & 0 - 25 Hz & 150 & 500 & 20 min & 20.5 min\\ \hline L\"ubeck\cite{munter2021BOE}& 600 kHz & 0 - 12 Hz & 100 & 500 & 41.6 s & 42.6 s
	\end{tabular}
	\label{tab:OurMethodVsLubeck}
\end{table}

\section{Discussion}
\subsection{Dynamic contrast source of three-domain structure in MCF-7 spheroids}
In Section \ref{Sec:ResultImageContrast}, AS-DOCT reveals three-domain structures in untreated MCF-7 spheroids.
To examine these regions, we compared AS-DOCT against authentic LIV (aLIV), Swiftness \cite{morishitaBOE2026}, and bivariate color-fusion representations \cite{morishita2025spie}.
Because aLIV and Swiftness measure local scatterer occupancy and motion speed, respectively, they aid in functional interpretation.
Furthermore, color-fusion imaging facilitates the visualization of multiple functional domains.

\begin{figure}
	\centering\includegraphics[width=13cm]{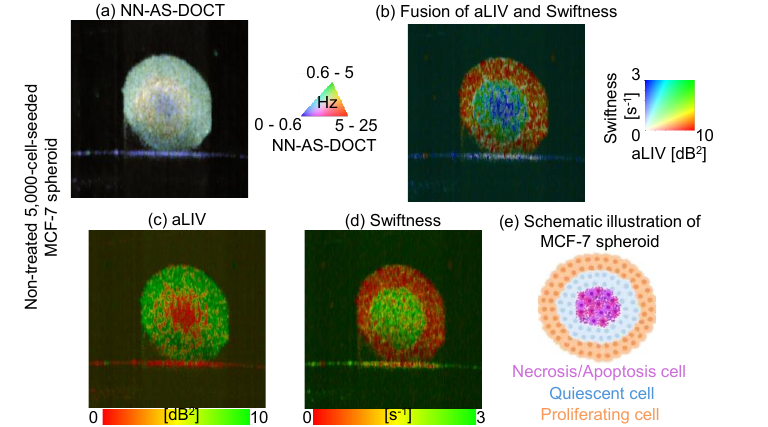}
	\caption{
		(a) NN-generated AS-DOCT of a 5,000-cell MCF-7 spheroid highlighting three domains, (b) aLIV/Swiftness color-fusion image, (c) aLIV image, (d) Swiftness image, and (e) schematic model of an MCF-7 spheroid (reprinted from Ref.\@\cite{malhao2022Toxics}).
	}
	\label{Fig:InterpretationMcf7}
\end{figure}
The aLIV and Swiftness color-fusion image [Fig.\@ \ref{Fig:InterpretationMcf7}(b)] resolves three domains matching those in NN-AS-DOCT [Fig.\@ \ref{Fig:InterpretationMcf7}(a)].
The core domain exhibits low aLIV and high Swiftness signals [Fig.\@ \ref{Fig:InterpretationMcf7}(c) and (d)], indicating that a small fraction of scatterers move rapidly.
The intermediate domain displays high aLIV and high Swiftness, indicating a high occupancy of dynamic scatterers with fast motion.
In contrast, the periphery shows high aLIV and low Swiftness, indicating a high occupancy of slow-moving scatterers.

It is known that, in MCF-7 spheroids, nutrient gradients drive the formation of three regions (Fig.\@ \ref{Fig:InterpretationMcf7}(e), adapted from Ref.\@ \cite{malhao2022Toxics}); a necrotic/apoptotic core, a proliferating periphery, and a quiescent intermediate zone.

This alignment suggests that NN-AS-DOCT effectively maps these distinct physiological cell states.

\subsection{Limitations}\label{Sec:Limitation}
Our method has two main limitations.
First, the quantitative accuracy of NN-derived AS-DOCT metrics remains constrained.
As noted in Section \ref{Sec:AS-DOCT-metrics}, agreement between NN-predicted and ground-truth values declines within the necrotic core for middle- and high-frequency channels.
Because AS-DOCT is primarily used for qualitative visualization, this limitation does not hinder visual interpretation.
However, caution is advised when extracting quantitative metrics from these regions.

Second, model generalization across different hardware systems and biological tissues remains unverified.
The generalization of the model by including diverse biological specimens and alternative OCT configurations in the training dataset represents an important direction for future work.

\section{Conclusion}
Our proposed NN framework, integrated with a multi-burst scanning protocol, generates high-quality AS-DOCT images from only 16 OCT frames.
The NN-generated AS-DOCT resolved multi-domain structures across three cancer spheroid types, matching conventional 512-frame reconstructions.
Furthermore, the approach enabled fast volumetric AS-DOCT, capturing full 3D volumes in 26.2 s.

In conclusion, NN-based AS-DOCT accelerates 3D dynamic imaging, providing a feasible path toward high-throughput, image-based screening applications.

\section*{Funding}
Japan Science and Technology Agency (JPMJCR2105, JPMJSP2124); Japan Society for the Promotion of Science (26H02451, 24KJ0510, 22KF0058, 22K04962, 21H01836).

\section*{Acknowledgment}
Yusong Liu is supported by the Japan Science and Technology Agency through the SPRING scholarship program.
Rion Morishita is supported by the Japan Society for the Promotion of Science through the Research Fellowship for Young Scientists.

\section*{Disclosures}
Liu, El-Sadek, Morishita, Yasuno: Nidek (F), Sky Technology (F), Panasonic (F), Nikon (F), Santec (F), Kao Corp. (F), Topcon (F).
Furukawa, Matsusaka: None.

\section*{Data Availability Statement}
Data underlying the results presented in this paper are not publicly available at this time but maybe obtained from the authors upon reasonable request.
\section*{Supplementary document}
The supplementary document presents images of all evaluation samples.
\bibliography{Yusong_AsDoct}

\pagebreak

\setcounter{figure}{0}
\renewcommand\thefigure{S\arabic{figure}}   
\setcounter{table}{0}
\renewcommand\thetable{S\arabic{table}}   
\setcounter{section}{0}
\renewcommand\thesection{S\arabic{section}}  

\section*{\Large Supplementary figures}
\begin{figure}[h]
	\centering\includegraphics[width=13cm]{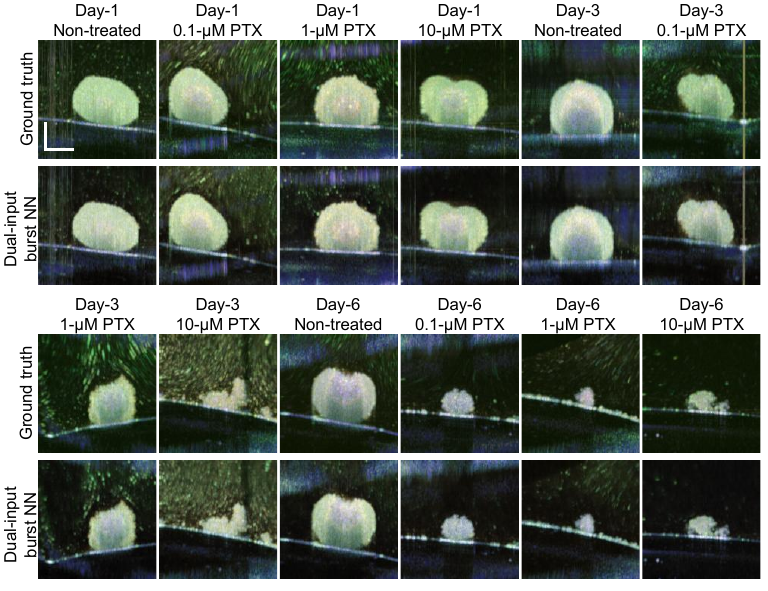} 
	\caption{%
		The cross-sectional AS-DOCT of the ground truth and NN (dual-input burst) generated versions.
		The samples are 12 breast cancer (MCF-7) spheroids seeded by 1,000 cells, which were used for evaluation.
		This figure supplements Fig.\@ 3.
	}
	\label{SFig:1000CellMCF7}
\end{figure}

\begin{figure}[h]
	\centering\includegraphics[width=13cm]{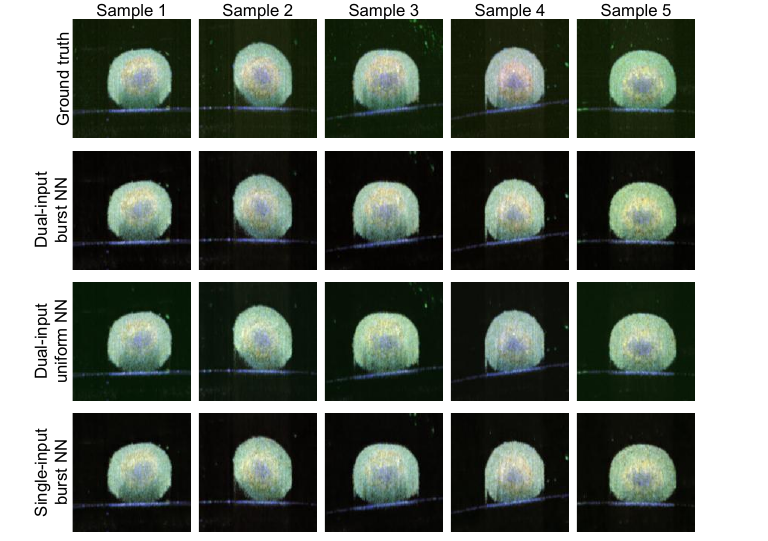} 
	\caption{%
		The performance comparison of three NN models with all the five 5,000-cell-seeded breast cancer (MFC-7) spheroids.
		The models being compared are dual-input burst, dual-input uniform, and single-input burst NN models.
		This figure supplements Fig.\@ 7.
	}
	\label{SFig:5000CellMCF7}
\end{figure}

\vspace{3ex}
These supplementary figures (Figs.\@ \@\ref{SFig:1000CellMCF7} to \@\ref{SFig:5000CellMCF7}) respectively supplement Fig.\@ 3, 4 and 7 by showing the results of other spheroid samples.

\clearpage

\begin{figure}[h]
	\centering\includegraphics[width=13cm]{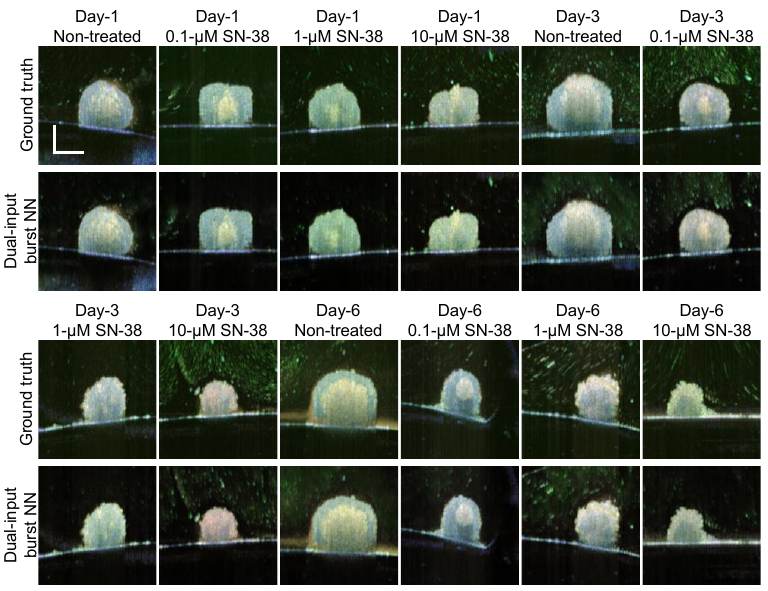} 
	\caption{%
		The cross-sectional AS-DOCT of the ground truth and NN (dual-input burst) generated versions.
		The samples are 12 colon cancer (HT-29) spheroids seeded by 1,000 cells, which were used for evaluation.
		This figure supplements Fig.\@ 4.
	}
	\label{SFig:1000CellHT29}
\end{figure}
\clearpage

\end{document}